\documentclass[10pt]{article}

\title{Shape Dynamics\footnote{Based on a talk given at Loops'11 (Madrid)}}
\author{\bf Tim Koslowski\footnote{\href{mailto:tkoslowski@perimeterinstitute.ca}{tkoslowski@perimeterinstitute.ca}}
\\\it Perimeter Institute for Theoretical Physics\\\it 31 Caroline Street, Waterloo, Ontario N2L 2Y5, Canada}

\usepackage{amssymb}
\usepackage{amsmath}
\usepackage[usenames,dvipsnames]{color}
\usepackage{hyperref}

\usepackage[left=1in,right=1in,top=1in,bottom=1in]{geometry}                  

\hypersetup{
    colorlinks=true,
    linkcolor=blue,
    citecolor=red, 
    urlcolor=Violet}

\begin{document}

\maketitle

\begin{abstract}
  General Relativity can be reformulated as a geometrodynamical theory, called Shape Dynamics, that is not based on spacetime (in particular refoliation) symmetry but on spatial diffeomorphism and local spatial conformal symmetry. This leads to a constraint algebra that is (unlike General Relativity) a Lie algebra, where all local constraints are linear in momenta and may thus be quantized as vector fields on the geometrodynamic configuration space. The Hamiltonian of Shape Dynamics is complicated but admits simple expressions whenever spatial derivatives are negligible.
\end{abstract}

Shape Dynamics \cite{ShapeDynamics} (for motivations see \cite{PreShapeDynamics}) is a canonical geometrodynamic theory that is equivalent General Relativity, but with spacetime refoliation invariance traded for local spatial conformal invariance. In other words: there are two different theory spaces that contain the dynamics of gravity, both defined on the same metric field content but with two distinct symmetry principles. Knowing the theory space is important because detail of a theory space can alter properties of effective field theories, in particular nonperturbative renormizability. 

\subsubsection*{Spacetime Versus Spatial Symmetry}

Abandoning spacetime refoliation invariance for a spatial symmetry is not just a change of symmetry, but also abandons the beautiful Hamiltonian construction principle for General Relativity of Hojman, Kucha\v{r} and Teitelboim \cite{HojmanKucharTeitelboim}. The starting point of their work is spacetime diffeomorphism symmetry and in particular its generators, the Lie-algebra of spacetime vector fields. These yield the Dirac-algebra of hypersurface deformations when pulled back to a Riemannian hypersurface:
\begin{equation}\label{equ:DiracAlgebra}
 [H(u),H(v)]=H([u,v]]),\,\,\,[H(u),S(N)]=S(\mathcal L_u N),\,\,\,[S(N),S(M)]=H(N\nabla M-M\nabla N),
\end{equation}
where spacetime vectors have been decomposed into spatial parts $u$ and $v$ tangent to the hypersurface and refoliation parts $N$ and $M$. Then \cite{HojmanKucharTeitelboim} proves a representation theorem of the Dirac algebra on the phase space of geometrodynamics, where the fundamental canonical pair is the spatial metric $g_{ab}(x)$ and its canonically conjugate momentum density $\pi^{ab}(x)$. Assuming time reversibility and a strong notion of locality, one finds that the generators $H(u)$ and $S(N)$ have to take the form of the ADM constraints
\begin{equation}\label{equ:ADMconstraints}
 H(u)=\int d^3x \pi^{ab}\mathcal L_u g_{ab},\,\,\,S(N)=\int d^3 x N \left(\frac{1}{\sqrt{|g|}}\pi^{ab}G_{abcd}\pi^{cd}-(R-2\Lambda)\sqrt{|g|}\right),
\end{equation}
where $G_{abcd}$ denotes the DeWitt supermetric. This is precisely the form of constraints that one obtains from the Legendre transform of the Einstein--Hilbert action. In other words one finds the canonical form of the Einstein--Hilbert action directly from spatial diffeomorphism and refoliation symmetry. The price of spacetime symmetry however is the complicated bracket $[S(N),S(M)]=H(N\nabla M-M\nabla N)$ which involves structure functions and a representation of $S(N)$ that is non-linear in the momenta. Both complications are associated with refoliation invariance and are thus eliminated if refoliation invariance is traded for spatial conformal invariance, whose generator is linear in the momenta and whose algebra is simple.

\subsubsection*{Canonical Best Matching}

Canonical best matching, which is a Hamiltonian realization of Barbour's Lagrangian best matching, is a construction that implements a configuration space symmetry in a dynamical system. Consider a dynamical system $(\Gamma,\{.,.\},\{\chi^\mu\}_{\mu\in\mathcal M})$ on phase space $\Gamma$ with Poisson bracket $\{.,.\}$ and first class constraints\footnote{We include the Hamiltonian in the set of first class constraints as an energy conservation constraint.} $\{\chi_\mu\}$ and denote the configuration variables by $\{q_i\}_{i\in \mathcal I}$. We want to implement the action of a symmetry group $\mathbb G$ acting as $ q_i \mapsto Q_i(q,\phi)$, where we choose the group parameters $\{\phi_\alpha\}_{\alpha \in \mathcal A}$ such that $\phi_\alpha=0$ denotes the unit element. The canonical best matching procedure consists of three steps:\\
1. Extend the original phase space $\Gamma$ by $T^*(\mathbb G)$ equipped with the canonical symplectic structure\footnote{For simplicity we consider only commutative groups at this point, but a similar treatment holds more generally.}, denote the momenta canonically conjugate to $\phi_\alpha$ by $\pi^\alpha$. We retain the original system by adjoining to the set of first class constraints the additional constraints $\{\pi^\alpha\}_{\alpha \in \mathcal A}$.\\
2. Implement $\mathbb G$-symmetry through a canonical transformation $T$ generated by $F=Q_i(q,\phi)P^i+\phi_\alpha \Pi^\alpha$, where capitals denote the transformed variables and $p^i$ is canonically conjugate to $q_i$. This turns the first class constraints $\pi^\alpha$ into a conserved charge that commutes with all $T\chi_\mu$
\begin{equation}
 T:\pi^\alpha \mapsto \pi^\alpha-\pi_i^\alpha(q)p^i. 
\end{equation}
3. Impose the ``best matching'' (or ``pure gauge'') condition $\pi^\alpha=0$. This may lead to three different types of dynamical systems:\\
(a) $\pi^\alpha$ is first class. This means that the system was already $\mathbb G$-symmetric.\\
(b) $\pi^\alpha$ gauge-fixes (some of) the $\chi_\mu$. We will see in a moment that this allows us to trade symmetries.\\
(c) $\pi^\alpha$ generates secondary constraints; in this case one has to follow the Dirac procedure to analyze the system. No general statements about the system are possible in this case.\\
The generic case is of course a mixture of the three prototype cases (a-c).

\subsubsection*{Linking Gauge Theories}

Let us now suppose canonical best matching yields case (b). Then we define a ``Linking Gauge Theory`` on $\Gamma\times T^*(\mathbb G)$ with first class constraints $T \chi_\mu,T \pi^\alpha$. The constraints can be split into three disjoint sets: \\
(1) The constraints $T\chi_\mu$ that are gauge fixed by $\pi^\alpha$. These are locally solvable for $\phi_\alpha$ by the implicit function theorem, so they are of the form $\chi^1_\alpha=\phi_\alpha-\phi_\alpha^o(q,p)$.\\
(2) The constraints $T\pi^\alpha$, which are of the form $\chi_2^\alpha=\pi^\alpha-\pi_i^\alpha(q)p^i$.\\
(3) The rest of the $T\chi_\mu$, which can be shown to not be invertible for $\phi_\alpha$, i.e. these constraints $\chi^3_\nu$ weakly commute with the $\pi^\alpha$ because we assumed best matching yields a theory of type (b).\\
We can now implement two partial gauge-fixing conditions for this linking theory:

The first is $\phi_\alpha=0$, which gauge fixes only the $\chi_2^\alpha$ constraints, whose Lagrange multipliers are set to vanish. We can thus perform phase space reduction by setting $(\phi_\alpha,\pi^\alpha)=(0,\pi_i^\alpha(q)p^i)$ and removing the constraints $\chi_2^\alpha$. Using $q_i=Q_i(q,0)$ and the fact that the $\pi^\alpha$ do not occur in the other constraints, we find that the constraints reduce to the set $\chi_\mu$ after phase space reduction, the reduced phase space coincides with $\Gamma$ and the Dirac bracket coincides with the Poisson bracket on reduced phase space, so we have recovered the gauge theory before best matching.

The second partial gauge fixing condition is $\pi^\alpha=0$. In this case the constraints $T\chi^1_\alpha$ drop out and the phase space reduction $(\phi_\alpha,\pi^\alpha)=(\phi^o_\alpha,0)$ again yields $\Gamma$ as the reduced phase space and the Dirac bracket again coincides with the Poisson bracket. The remaining constraints are not necessarily independent of $\phi_\alpha$, but because the $T\chi^3_\mu$ can not be solved for $\phi_\alpha$ they are weakly equivalent to $\chi^3_\mu$ and $\pi_i^\alpha(q)p^i$.

We can directly check that the two partially gauge-fixed theories are equivalent by imposing the $\chi^1_\alpha$ and $\chi_2^\alpha$ respectively as additional partial gauge fixing conditions, so the two theories have the same initial value problem and the same equations of motion. We have thus traded the constraints $\chi^1_\alpha$ of the first theory for the constraints $\pi_i^\alpha(q)p^i$ of the second theory without changing the phase space or Poisson bracket. If the two sets of constraints generate two different sets of gauge symmetries, then we have traded one set of gauge symmetries for another.

\subsubsection*{Pure Shape Dynamics}

Let us now best-match General Relativity (on a compact Cauchy surface without boundary) with spatial conformal transformations that preserve the total spatial volume. We introduce a conformal factor $\phi$ with canonically conjugate momentum density $\pi_\phi$ and apply the canonical transformation $T$ generated by
\begin{equation}
 F=\int d^3 x\left(g_{ab}e^{4\hat \phi}\Pi^{ab}+\phi\Pi_\phi\right)
\end{equation}
to ADM gravity (equation \ref{equ:ADMconstraints}). Here $g_{ab}$ denotes the spatial metric, $\pi^{ab}$ its canonically conjugate momentum density, capitals denote transformed quantities, and we have used the volume preserving conformal factor $\hat \phi(x)=\phi(x)-\frac 1 6 \ln\langle e^{6\phi}\rangle$ and the mean $\langle f\rangle=\frac{\int d^3x \sqrt{|g|}\,f\,}{\int d^3y \sqrt{|g|}}$. This yields the conserved charge $Q(x)=\pi_\phi(x)-4(\pi(x)-\langle \pi \rangle\sqrt{|g|(x)})$ in terms of $\pi=g_{ab}\pi^{ab}$ and $\langle\pi\rangle=\frac{\int d^3x \pi}{\int d^3y \sqrt{|g|}}$. It turns out that imposing $\pi_\phi(x)=0$ yields a theory of type (b), so we have a linking theory between General Relativity and a theory which we call Shape Dynamics. The conserved charge $Q(x)$ becomes a spatial conformal constraint $C(x)=\pi(x)-\langle \pi\rangle\sqrt{|g|(x)}$ that preserves the total spatial volume, while the ADM-diffeomorphisms $H(u)$ remain unchanged. All but one of th ADM scalar constraints are gauge fixed; the remaining one becomes the Shape Dynamics Hamiltonian $H_{SD}$. $H_{SD}$ is defined by the normalization $\langle e^{6\hat \phi}\rangle=1$ and
\begin{equation}\label{equ:LY}
  T S(x)=H_{SD}\sqrt{|g|(x)}e^{6\hat\phi(x)},
\end{equation}
which simplifies on the constraint surface $C(x)=0$ to a modified Lichnerowicz--York equation, which can be proven to have unique solutions $\hat\phi(x)$. The total Shape Dynamics Hamiltonian is 
\begin{equation}
 {\bf H}=\mathcal N H_{SD}+H(u)+C(\rho),
\end{equation}
where $\mathcal N$ is a single global Lagrange multiplier and $\rho(x)$ a local Lagrange multiplier. We see that all local constraints are linear in momenta at the expense of a nonlocal Hamiltonian $H_{SD}$.

The Lichnerowicz--York equation (\ref{equ:LY}) obstructs a straightforward derivation of $H_{SD}$, although one can prove the existence of $H_{SD}$. However, this is {\bf not} more complicated than General Relativity, where the analogous equation has to be solved in the initial value problem. Despite this, it is clear that the nonlocal Hamiltonian is a challenge for Shape Dynamics, but this seems to be an acceptable price to pay for its advantages compared with General Relativity:
\\
1. All (local) constraints are linear in momenta and generate vector fields on the geometrodynamic configuration space.\\
2. The constraint algebra is a Lie algebra and closes in a much simpler way than General Relativity:
\begin{equation}
 \begin{array}{rclcrclcrcl}
   \{C(\rho_1),C(\rho_2)\}&=&0&&\{C(\rho),H_{SD}\}&=&0&&\{H(u),H_{SD}\}&=&0\\
   \{H(u),C(\rho)\}&=&C(\mathcal L_u\rho)&&\{H(u),H(v)\}&=&H([u,v]).
 \end{array}
\end{equation}
3. Shape Dynamics is contained in a different theory space than General Relativity, which may change nonperturbative renormalizability.

The usual construction of effective field theories uses a theory space (i.e. field content, symmetries and locality) as well as dimensional analysis. We adopt $V$ and $\langle \pi\rangle$ in addition to local variables as SD-local. We can then mimic the usual construction of effective Hamiltonians: Write down all low-dimensional SD-local field monomials in the linking theory that are invariant under diffeomorphisms and banal transformations (i.e. invariant under $g_{ab}\to\left.Tg_{ab}\right|_{\phi=\sigma},\pi^{ab}\to \left.T\pi^{ab}\right|_{\phi=\sigma},\phi\to\phi-\sigma$) that preserve the total spatial volume. Then determine the conformal field $\phi$ as a function of $g_{ab},\pi^{ab}$ by requiring local spatial conformal invariance.

\subsubsection*{Explicit Shape Dynamics}

There are at least three instances where the Shape Dynamics Hamiltonian can be explicitly constructed:\\
1. Using the fact that the solution to (\ref{equ:LY}) is homogeneous if all coefficient functions are homogeneous and that the normalization condition implies that $\hat \phi$ vanishes if $\phi$ is homogeneous, one finds the Shape Dynamics Hamiltonian to coincide with the temporal gauge Hamiltonian $H(N\equiv 1)$ on the restricted phase space 
\begin{equation}
 \Gamma_r=\left\{(g_{ab},\pi^{ab})\in \Gamma_{ADM}: \frac{\sigma^{ab}\sigma_{ab}}{|g|}=\langle\frac{\sigma^{ab}\sigma_{ab}}{|g|}\rangle,R=\langle R\rangle\right\},
\end{equation}
where $\sigma^{ab}=\pi^{ab}-\frac 1 3 \pi g^{ab}$. Notice that we can use $R=\langle R\rangle$ as a gauge fixing for local conformal symmetry, so $\Gamma_r$ contains {\it{three out of four}} local degrees of freedom of Shape Dynamics.\\
2. In the strong gravity (BKL) limit, one can neglect the spatial derivatives, so (\ref{equ:LY}) becomes an algebraic equation which can be solved giving the nonlocal Hamiltonian
\begin{equation}
 H_{SD}^{BKL}=V\left(\langle\sqrt{\sigma^{ab}\sigma_{ab}}\rangle^2-\frac 1 6 \langle\pi\rangle^2+2\Lambda\right).
\end{equation}
This may be a good starting point for quantization, because in light of the BKL-conjecture \cite{BKL} one expects the effects of quantum gravity to be most important when this Hamiltonian becomes exact.\\
3. In 2+1 dimensions on the sphere and torus one can solve (\ref{equ:LY}) explicitly, because $\Gamma_r$ contains all physical degrees of freedom in this case. The Shape Dynamics Hamiltonian turns out to coincide with the reduced phase space Hamiltonian for these topologies and one can thus explicitly Dirac quantize Shape Dynamics in this case. 

\subsubsection*{Summary}

1. Shape Dynamics is a geometrodynamic theory that is equivalent to General Relativity, but refoliation invariance is traded for local spatial conformal invariance.\\
2. Linking Theories (and canonical best matching) provide a general mechanism for trading gauge symmetries.\\
3. The Shape Dynamics Hamiltonian exhibits nonlocality for $1/4$ of the physical degrees of freedom.\\
4. There exists a construction principle for effective Hamiltonians based on the symmetries of Shape Dynamics.\\
5. There are instances where Shape Dynamics can be explicitly constructed (in particular: strong gravity, 2+1 torus and sphere).\\
6. Since all local constraints are linear in momenta, one can perform Dirac quantization on the 2+1 torus and investigate the large-CMC-volume/CFT-correspondence.\\
{\bf Acknowledgments:} I thank J. Barbour for helpful comments on the draft. Research at the Perimeter Institute is supported in part by the Government of Canada through NSERC and by the Province of Ontario through MEDT.

\bibliographystyle{utphys}

\end{document}